\documentclass[preprint,superscriptaddress,showpacs]{revtex4}
\usepackage[dvips]{graphicx}
\usepackage{color}
\usepackage{multirow}

\begin{document}

\title{
Rigidity of the conductance of anchored dithioazobenzene opto-mechanical switch
}

\author{M. Zemanov{\' a} Die{\v s}kov{\' a}}
\affiliation{
Department of Physics, 
Faculty of Electrical Engineering and Information Technology,
Slovak University of Technology, 
Ilkovi\v{c}ova 3, 812 19 Bratislava, Slovak Republic
}

\author{I. {\v S}tich}
\affiliation{
Institute of Physics, CCMS, Slovak Academy of Sciences, 84511~Bratislava, Slovakia
}

\author{P. Bokes}
\email{peter.bokes@stuba.sk}
\affiliation{
Department of Physics, 
Faculty of Electrical Engineering and Information Technology,
Slovak University of Technology, 
Ilkovi\v{c}ova 3, 812 19 Bratislava, Slovak Republic
}
\affiliation{European Theoretical Spectroscopical Facility (ETSF)}

\date{\today}

\begin{abstract}
Reversible opto-mechanical molecular switch based on a single azobenzene molecule suspended via thiolate links between realistic models of gold tips is investigated. Using a combination of the transfer matrix technique and density functional theory we focus on conductance of the nano-device in the two (meta)stable \textit{cis} and \textit{trans} junction conformations. We find the conductance of both conformations to be broadly similar. In qualitative agreement with related experiments, we find that the same nano-device with one/two methylene linker group(s) inserted on one/both ends of the azobenzene molecule is driven into tunneling regime and reduces the conductances by up to two orders of magnitude, again almost uniformly for both conformations. These results clarify the huge differences in switching ratios found previously and indicate that this nano-device is not particularly suited for use as a molecular switch based on conductance change.
\end{abstract}

\pacs{                                                                
85.65.+h 
73.63.-b 
73.22.-f 
}

\maketitle

\section{\label{sec:intro}Introduction}

Molecular electronics is undoubtedly a promising complementary technology to
semiconductor-based electronics~\cite{tans_97,reed_97,joachim_95,moresco_95,emberly_03,li_04,zhang_04,zhang_06,turansky_10_1,turansky_10_2}. 
The molecule used to fabricate a nano-scale device must be capable of rectification, amplification, or switching. 
Here we deal with the latter functionality, 
which is perhaps also the most prominent among these functionalities of single-molecule devices.
In order for a molecule to be used in a molecular switch it must have two states with very different properties, such as, for instance, conductance. This is typically achieved by modifying the molecular conformation. Change of the conformation may be driven by electric currents~\cite{qiu_04,choi_06,milosevic_07}, 
tip of an AFM/STM apparatus~\cite{moresco_95,emberly_03,Comstock2005,Henningsen2008,konopka_08,turansky_10_1,turansky_10_2,Quek2009}, 
or by electromagnetic radiation~\cite{li_04,zhang_04,zhang_06,turansky_10_1,turansky_10_2,balzani_00,feringa_00,collin_98}. The primary advantage of light-driven optoelectronic devices is their fast switching speed~\cite{li_04,zhang_04,zhang_06,turansky_10_1,turansky_10_2}. However, use of these optoelectronic devices requires that their operation be reversible. While there are numerous molecules capable of photo-switching in gas-phase, their embedding in junction by anchoring to tips or surfaces may compromise their switching capability either due to mechanical hindrance~\cite{turansky_10_1,turansky_10_2} or quenching of excited state in the photo-switchable process~\cite{dulic_03}.

One of the simplest photo-switchable molecules considered for optoelectronic applications is azobenzene (AB, C$_{12}$H$_{10}$N$_{2}$). AB has a relatively simple molecular structure with \textit{cis} and the energetically more stable \textit{trans} isomers, see Fig.~\ref{fig:mechano_switch}. The length of \textit{trans} isomer is markedly longer compared to the \textit{cis} isomer by over 2~\AA~and, based on theoretical modeling, the conductance of the \textit{trans} isomer was reported to be two orders of magnitude larger than that of its \textit{cis} counterpart~\cite{zhang_04,zhang_06}. If the switching ratio (SR) is indeed as high, this nano-device would be an ideal incarnation of a molecular switch, provided the switch could be operated by photo-switching. However, high SRs are typically achieved by opening/closing of $\pi$-conjugated rings~\cite{kim2012_2}, which is not the case in AB. Moreover, up to now, low-temperature charge transport experiments using single-molecule junctions are scarce~\cite{kim2012_2,kim2012_1}. Hence, direct experimental verification could be provided only recently~\cite{kim2012_1}. The experiments for azobenzene-thiomethyl (ABTM) molecule find the \textit{cis} conformer more conducting than \textit{trans} and suggest only a small SR, with \textit{cis}/\textit{trans} conductances of 
(4.9$\pm$3.4)/(1.6$\pm$0.7)~$G_{0}$. 
The small SR is attributed to the presence of -CH$_{2}$- side chains acting as tunneling barriers (TB)~\cite{kim2012_1}. However, the same ABTM molecule was also considered in the theoretical treatment~\cite{zhang_04,zhang_06}, and hence, this does not explain the discrepancies between theory and experiment. To make things even more involved, recent theoretical modeling of self-assembled ABTM junctions on Au(111) surfaces~\cite{wang2012} yielded higher \textit{trans} than \textit{cis} conductance, fairly small SR, and transmissions significantly different from those corresponding to mechanically controlled break-junction (MCBJ)~\cite{afm-mcb-rev} type of treatment~\cite{kim2012_1}, suggesting that also coupling to electrodes may play an important role.    
In addition, other studies explored contacting the AB molecule with silicon~\cite{Nozaki09} or carbon nanotube~\cite{Valle2007} and found even larger SRs of the \textit{cis}/\textit{trans} conductance than those reported in Refs.~\cite{zhang_04,zhang_06}.
Additional question-marks mar the photo-switchability of anchored AB photo-switches. While anchored polymeric chains of AB molecules are easily switchable optically~\cite{hugel_02}, switching of their tip/surface-anchored single AB counterpart was found to be mechanically hindered~\cite{turansky_10_1,turansky_10_2}. In order to eliminate the mechanical hindrance a variant of an anchored single-molecule AB switch using a conducting carbon nano-tube support for one electrode was proposed~\cite{zhang_04}. 

The host of theoretical modeling done in different junction geometries but using essentially the same theoretical treatment yield vastly different results~\cite{zhang_04,zhang_06,wang2012,kim2012_1}. The only experiment, also using a particular junction geometry, finds virtually identical conductances for the two junction conformation, albeit with very large error bars. This situation invites a new theoretical study which would sort out the questions, preferably using a different formulation from the previous localized basis set approach~\cite{zhang_04,zhang_06,wang2012,kim2012_1}. To this end we focus here on the conductance of the \textit{trans} and \textit{cis} conformers of embedded 4,4$^{\prime}$-dithioazobenzene (DAB, C$_{12}$H$_{8}$N$_{2}$S$_{2}$), i.e. AB embedded in a gold-AB-gold junction via thiolate or thiomethyl linkers. In order to avoid the complexities of optical switching of the device, we study here a mechano-switch controllable in a reversible way~\cite{turansky_10_2}, which corresponds to a device controlled by AFM/STM tip. Particular attention is paid to realistic Au electrodes which are modeled emulating MCBJ-type of treatment. Transport is modeled using the Landauer-Buttiker scattering theory~\cite{buttiker_86} in density functional theory (DFT) formulated in plane wave basis~\cite{pwcond}. Plane waves are deemed important for a better description of the electronic states delocalized over the metallic tips~\cite{Strange08}. This is at variance with all previous theoretical modeling~\cite{zhang_04,zhang_06,wang2012,kim2012_1}, which used localized basis sets
that were not specifically designed for treatment of nano-junctions or surfaces~\cite{Garcia-Gil2009}.

We find the conductance of both \textit{trans} and \textit{cis} DAB isomers broadly similar, differing typically at most by a factor of $\approx$2. However, we find similar conductance differences also for two different realizations of the same molecular conformation. We note in passing that similarly small on/off conductance differences were found also in other related systems~\cite{Quek2009}. Hence, as we argue below, the discrepancies we can account for stem from differences in treatment of the leads and their coupling to the AB molecule. Contrary, in good agreement with related experiments~\cite{danilov_08}, we find that separating the electrodes from AB by methylene groups which act as TBs, lowers the conductance by about two orders of magnitude. We conclude that Au-DAB/ABTM-Au junction is \textit{not} particularly well suited for use as an opto-mechanical molecular switch.

\section{\label{sec:simulations}Simulation details}

Ground-state calculations were performed using the plane-wave-based DFT code \texttt{Quantum espresso}~\cite{Espresso}. The energy cutoff of 
30 and 300 Ryd was used to expand the electronic orbitals and electronic density, respectively, in combination with the ultrasoft pseudopotentials~\cite{vaderbilt_90} for description of atomic core electrons. The k-point sampling of the supercell containing a slab of $3 \times 4$ Au(111) surface consisting of 6 ideal layers of bulk Au, 10 and 13 Au atoms mimicking the tips of the upper and lower electrodes, see Sect.~\ref{ssec:geometry}, and the DAB molecule with optional additional methylene groups, see 
Figs.~\ref{fig:geometry},~\ref{fig:geometry-Me}, was used with a $4 \times 3\times 2$ shifted Monhorst-Pack k-point grid~\cite{monkhorst_76}. Exchange and correlation effects were described with the PBE functional~\cite{pbe_1,pbe_2}.

Transport properties of the junctions were obtained using the transfer matrix method~\cite{Joon1999} implemented in the \texttt{PWCOND} code~\cite{pwcond} with plane-wave basis and ultrasoft pseudopotentials. Compared to localized basis sets, use of plane-wave basis results in a better description of the electronic states delocalized over the disordered Au tips~\cite{Strange08}. Furthermore, use of the periodic cell in the direction perpendicular to the current flow enhances the bulk character of the electrodes, which appears to be an issue here. Taking the \textit{trans} geometry as an example, going from the $2 \times 2$ to the $4 \times 4$ Monkhorst-Pack 
k-point sampling in the conductance calculation results in $10\%$ change of the conductance. 
The $6 \times 6$ grid, which gives changes in the conductance smaller than $1\%$, was taken for calculation 
of the presented results.
The dense k-point sampling is needed to correctly describe the scattering states in the metal electrodes. 
By explicit increase of the surface cell size we have checked that this behavior is not related to interference effects between the two junctions in the neighboring 
supercells. For this purpose the $3 \times 4$ Au(111) surface supercell turned out to be sufficient. In addition, upper and lower electrodes had both three (111) layers of bulk Au, i.e. one single complete sequence of the A-B-C stacking of the fcc crystalline structure.

The DFT-based approach is known to be a poor approximation for the exact linear-response conductance~\cite{Koentopp05,Ferretti05,Quek07,Toher08,Mera10}, particularly for systems with transport dominated by Coulomb blockade (CB)~\cite{Kurth10,Thygesen09,Stadler2008}. In our junctions with the DAB molecule directly bonded to the electrodes the DAB molecule is well contacted to the electrodes which drives the junction away from the CB regime and leads to reliable predictions~\cite{Zotti2010}. 
Contrary, considering the thiomethyl AB molecule. i.e. with -CH$_2$- linkers on both ends, the results need to be interpreted with care. We note, however, that the relative conductance change due to these linkers is in good agreement with experimental results for related junctions~\cite{danilov_08}, see Sect.~\ref{ssec:methylene}.

\section{\label{sec:results}Results}

\subsection{\label{ssec:geometry}Junction geometry}

\begin{figure}[!t]
\includegraphics[width=0.45\textwidth]{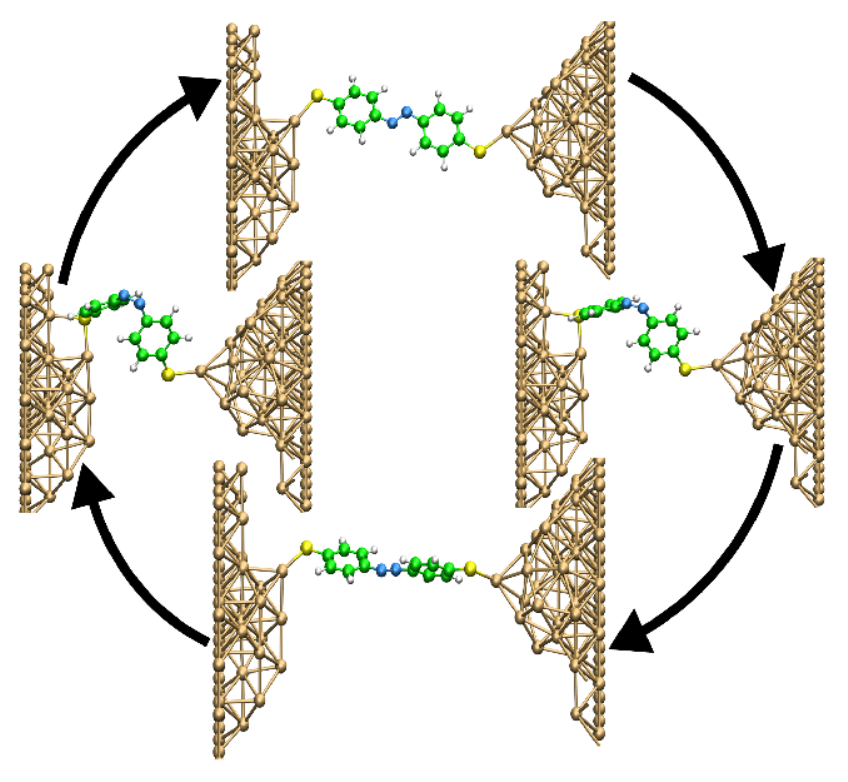}
\caption{
Mechanical switching of the model gold-DAB-gold junction with realistic gold tips featuring two complete 
\textit{trans}$\rightarrow$\textit{cis}$\rightarrow$\textit{trans} switching cycles~\cite{turansky_10_1,turansky_10_2}. Geometries of this mechano-switch were used to construct geometries for transport study, Fig.~\ref{fig:geometry}.
} 
\label{fig:mechano_switch}
\end{figure}

\begin{figure}[!h]
\includegraphics[width=0.45\textwidth]{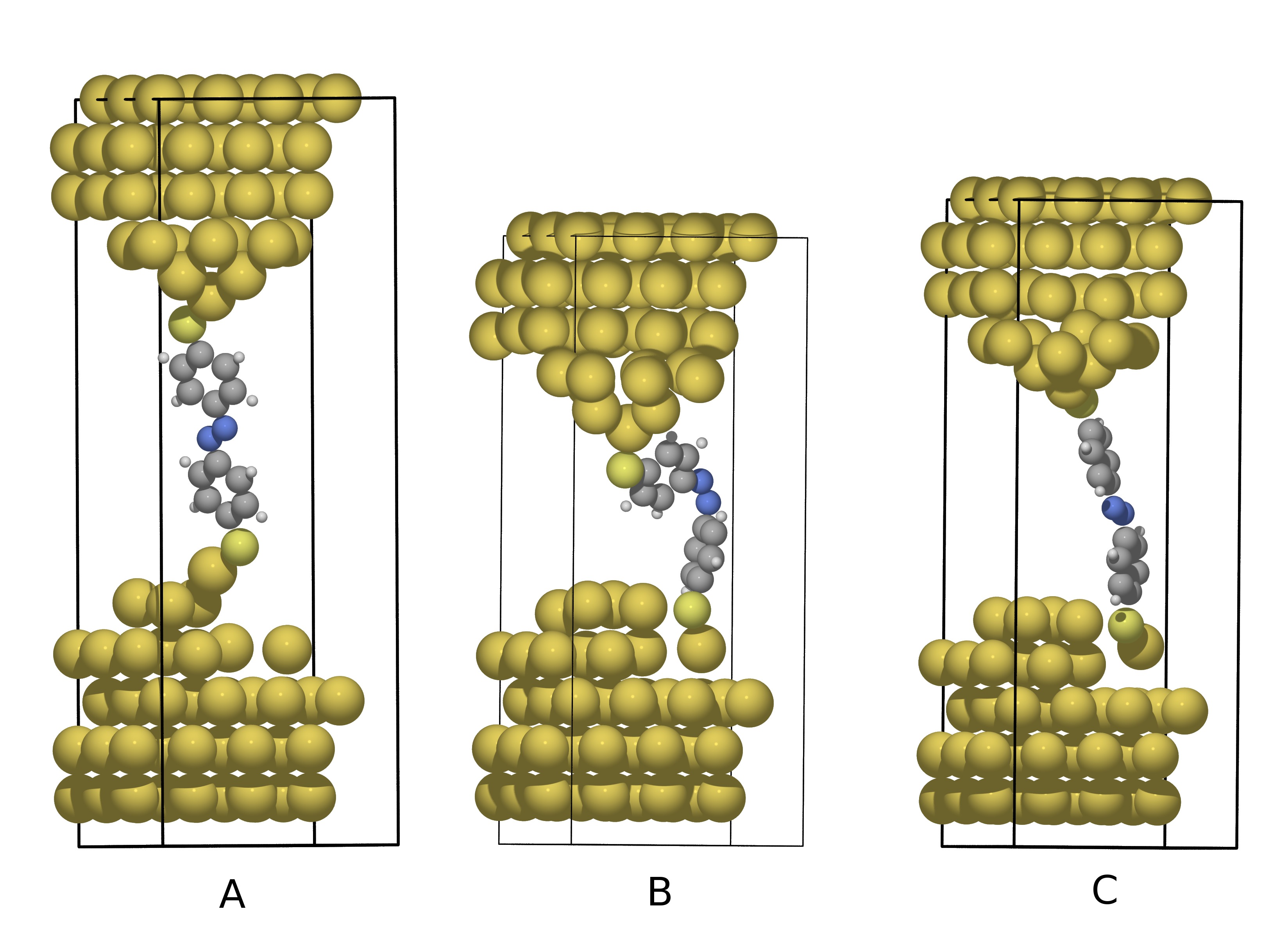}
\caption{
Geometries of the gold-DAB-gold junctions used to study electronic transport: 
A(\textit{trans}) [left], B(\textit{cis}) [center] and C(\textit{trans}) [right].
} 
\label{fig:geometry}
\end{figure}

The starting point of our modeling of the gold-DAB-gold junction were the low-temperature geometries of the junction in \textit{trans}/\textit{cis} conformations sampled from the trajectories previously used to study the opto-mechanical switching~\cite{turansky_10_1,turansky_10_2}. The gold tips were prepared in a MCBJ-type procedure using an empirical effective medium theory potential~\cite{emt} together with Langevin dynamics. A gold rod coupled to two gold plates with applied periodic boundary conditions was pulled apart to the point of breaking the Au nano-junction and then bridged by DAB molecule. The junction so prepared was then treated with DFT techniques emulating a mechano-switching cycle
which consisted of series of geometry optimizations for variable distance between selected bulk layers
in the left and right electrodes.
We have used a model junction capable of reversible mechanical switching, shown in Fig.~\ref{fig:mechano_switch}. We note that from the limited set of junctions studied about one third showed this capability. However, such a junction was found impractical for transport study as the electrodes prepared by a protocol emulating MCBJ-type of treatment~\cite{afm-mcb-rev} were unnecessary large. For that reason we have modified the relaxed metal tip geometries so that the upper/lower electrodes consisted of semi-infinite Au(111) $3 \times 4$ surfaces plus 10/13 gold atoms of the upper/lower contact region, respectively. The metal atoms emulating the terminations were kept fixed along with the DAB molecular structure, all constrained to the geometry found previously in a fully consistent treatment~\cite{turansky_10_1,turansky_10_2}. All other Au atoms in the electrode were allowed to relax so as to provide a good metallic coupling between the electrodes and the junction. The resulting junction structure corresponding to the \textit{trans} and \textit{cis} DAB conformations are shown in Fig.~\ref{fig:geometry}. The sensitivity of the calculated transmission to the number and position of these relaxed Au atoms was checked and found to have only a negligible effect, see Sect.~\ref{ssec:conductance}. We note that the two \textit{trans} geometries (A, C) in Fig.~\ref{fig:geometry} correspond to two configurations resulting from reversible cyclic mechanical switching (upper/lower \textit{trans} conformations in Fig.~\ref{fig:mechano_switch}). 

\begin{figure}[!h]
\includegraphics[width=0.45\textwidth]{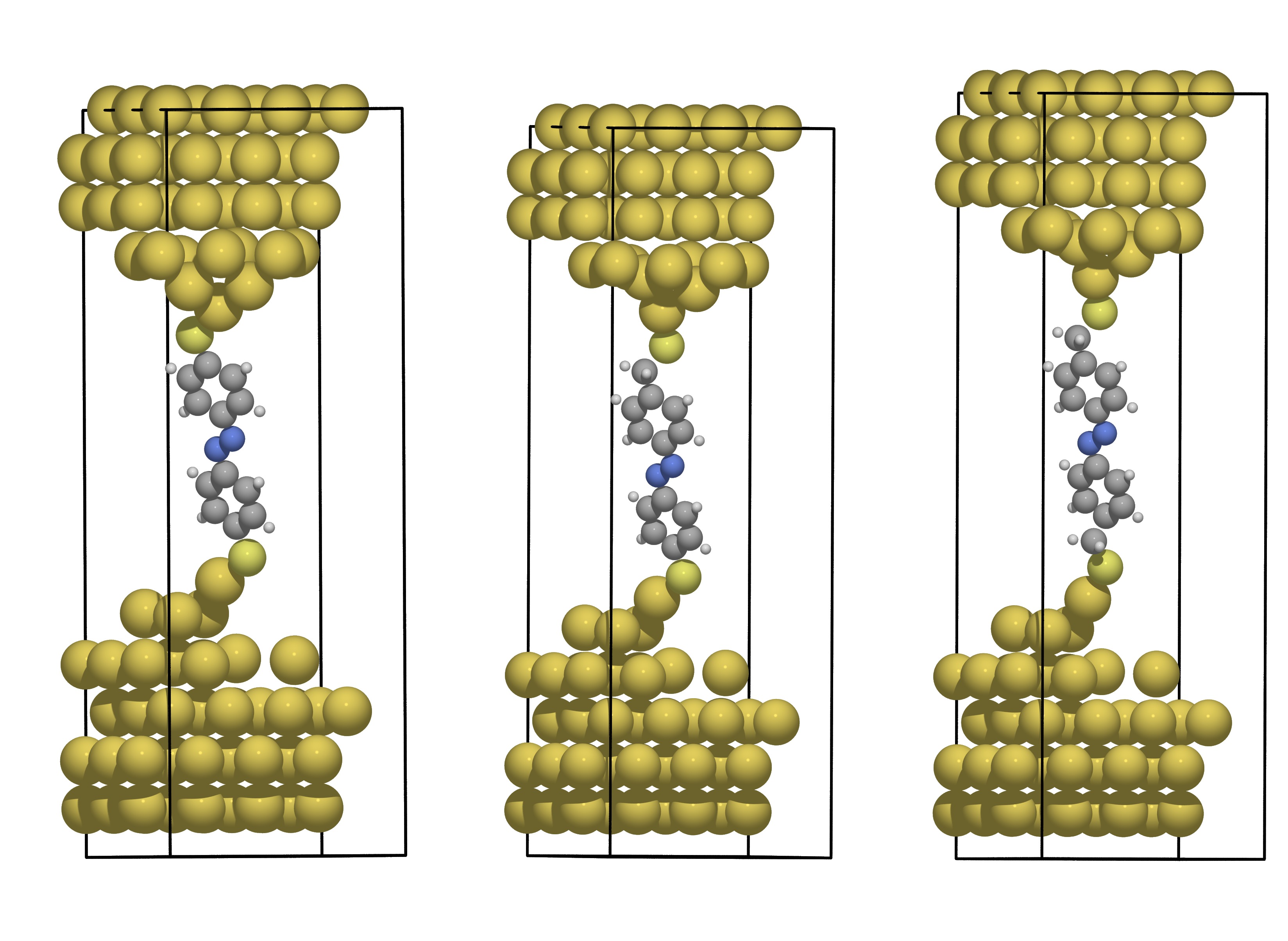}
\caption{
Models of the A(\textit{trans}) junctions with no [left], single [center]  and two [right] methylene groups 
modifying the contacts to the electrodes.
} 
\label{fig:geometry-Me}
\end{figure}

The \textit{cis} and \textit{trans} junctions with inserted methylene groups, Fig.~\ref{fig:geometry-Me}, were constructed by displacing one of the electrodes with all the Au and S atoms by a distance $\Delta z$ from the rest of the system and inserting the -CH$_2$- group into the resulting gap. The structure was then relaxed until the forces on the inserted group and the AB molecule dropped below $0.01$Ry/$a_B$. While this is a fairly large residual force, further optimization did not lead to any substantial change in the conductance, which is of primary concern here. For instance, for construction of the A(\textit{trans}) junction with additional -CH$_2$- group three values of $\Delta z$, 1.4~\AA, 1.5~\AA~and 1.6~\AA, were considered from which the geometry with the smallest total energy was selected for further 
conductance calculations.

\subsection{\label{ssec:conductance}Transmittance and conductance}

\begin{figure}[!h]
\begin{center}
\includegraphics[width=0.35\textwidth]{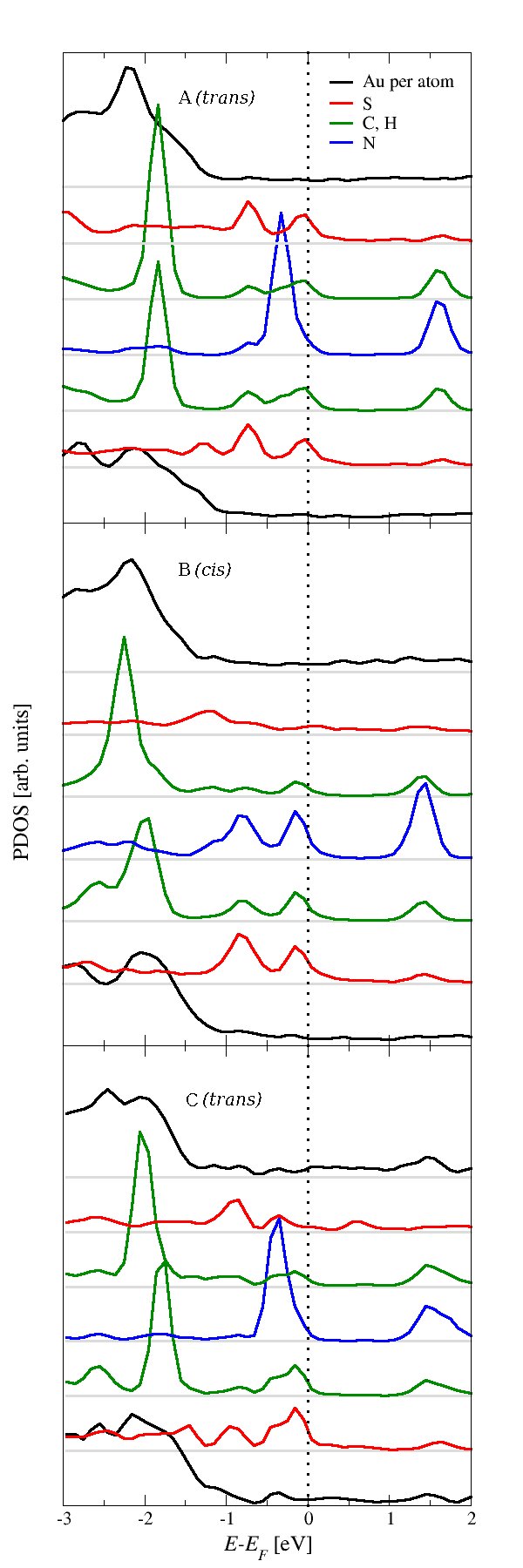}
\caption{
Projected density of states for the A(\textit{trans}) [top], B(\textit{cis}) [center] and C(\textit{trans}) [bottom] conformations of the contacted AB molecule. PDOS for the gold electrodes, sulfur linkers, and benzene rings are shown separately for the upper and lower parts of the junction. In all cases the Fermi energy is pinned to the upper tail of the HOMO resonance.
} 
\label{fig:PDOS}
\end{center}
\end{figure}

We start by analyzing results for DAB suspended in the junction between Au electrodes. The first indication of the rigidity of conductance of DAB upon reisomerization is visible already in the projected densities of states (PDOS) close to the Fermi energy. In Fig.~\ref{fig:PDOS} PDOS for three selected junction geometries are shown: A and C correspond to \textit{trans}-like conformations, whereas structure B is a \textit{cis}-like conformation. The biggest difference is found between the \textit{cis} and the \textit{trans} PDOS, signaled as a split of the \textit{cis} HOMO (Highest Occupied Molecular Orbital) level on the nitrogen atoms. This change, though, does only little to the offset of the Fermi energy from the HOMO level. This result is in agreement with general trends in the alignment of the HOMO orbital in $\pi$-conjugated thiolate molecules on gold surface~\cite{Heimel2006} and agrees qualitatively also with results for SAMs of ABTM on gold~\cite{wang2012}. Such a finding has important consequences for the conductance. 
Somewhat smaller differences are also found in PDOS corresponding to the two benzene moieties at energies around 2 eV below $E_{F}$. For the A(\textit{trans}) conformation both peaks are at the same energy, while those peaks are shifted to lower energies for the B(\textit{cis}) structure. On reisomerization to the C(\textit{trans}), one of those peaks remains aligned with the position corresponding to C(\textit{trans}), whereas the other returns back to the position corresponding to A(\textit{trans}). 

\begin{figure}[!h]
\includegraphics[width=0.5\textwidth]{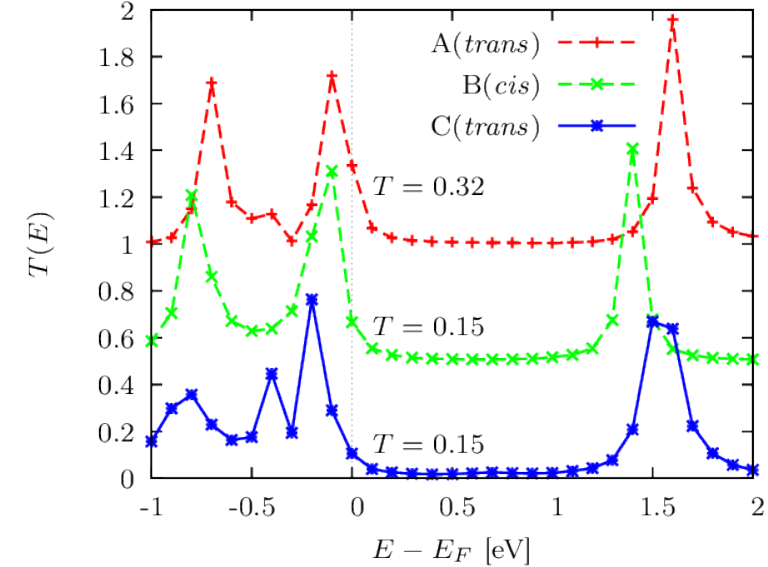}
\caption{
The transmissions for the A(\textit{trans}), B(\textit{cis}), and C(\textit{trans}) geometries (transmissions are shifted up by 0.5 for clarity). All the geometries give comparable values of the conductance (transmission at $E=E_F$) due to the pinning of the Fermi energy at the top of the HOMO resonance.
}
\label{fig:transmission}
\end{figure}

As demonstrated by Fig.~\ref{fig:transmission}, the observed rigidity of the HOMO offset from the Fermi energy is paralleled by a similar behavior of the junction transmission. Going from the A(\textit{trans}) to the B(\textit{cis}) conformation, the conductance of the junction decreases merely by a factor of $\approx$2.  
However, and most importantly, the other \textit{trans}-like geometry sampled from our mechano-switching cycle, C(\textit{trans}), gives conductance equal to that of 
B(\textit{cis}) conformation. Hence conductances of two \textit{trans}-like geometries differ also by a factor of $\approx$2. This indicates that the gross details of the junction geometry \textit{do} play a role in the resulting transport properties. Incidentally, a spread in the \textit{trans} conductances is in line with the large error bar experimentally observed for the \textit{trans} conductance~\cite{kim2012_1}.

\begin{figure}[!h]
\includegraphics[width=0.5\textwidth]{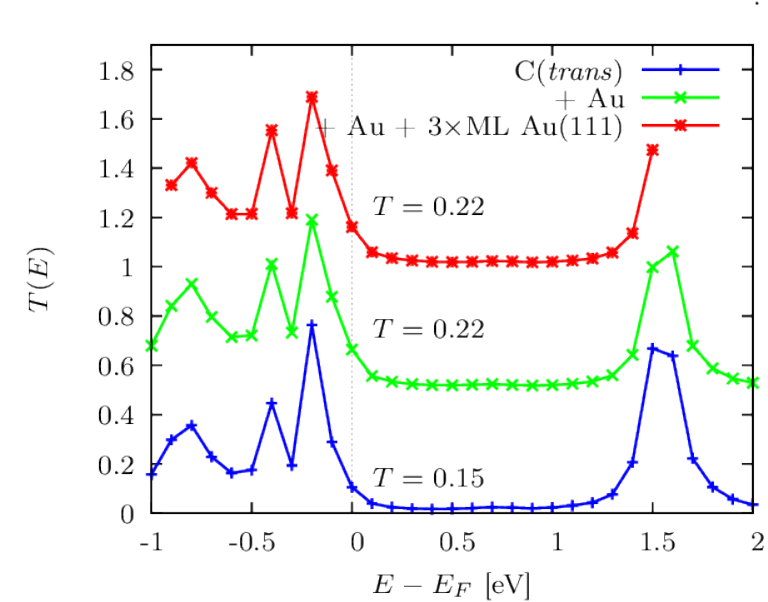}
\caption{
Rigidity of the transmissions of the C(\textit{trans}) junction geometry upon adding additional Au atom and additional layers of the Au(111) surface to the gold tip leading essentially to identical transmissions and conductances.
} 
\label{fig:check}
\end{figure}

In order to further explore the sensitivity of the transmission and conductance to finer details of the electrodes, we have slightly modified the details of the Au electrodes leaving the gross molecular features and coupling to electrodes unchanged. 
In particular in the C(\textit{trans}) junction an additional Au atom was placed into the top contact 
in Fig.~\ref{fig:geometry}C, occupying one of the left hollow sites of the Au(111) surface and bonding to some of the Au atoms of the tip,
and the number of layers in the bulk region of the gold tip was extended. From the resulting transmissions and conductances shown in Fig.~\ref{fig:check} we conclude that such gentle changes do \textit{not} play any important role. This reassures the rigidity of the conductance upon mechanical switching and suggests that the gold-DAB-gold junction is \textit{not} particularly suitable for use as an electronic molecular switch.

\subsection{\label{ssec:methylene}The effect of methylene groups on conductance}

\begin{figure}[!h]
\includegraphics[width=0.5\textwidth]{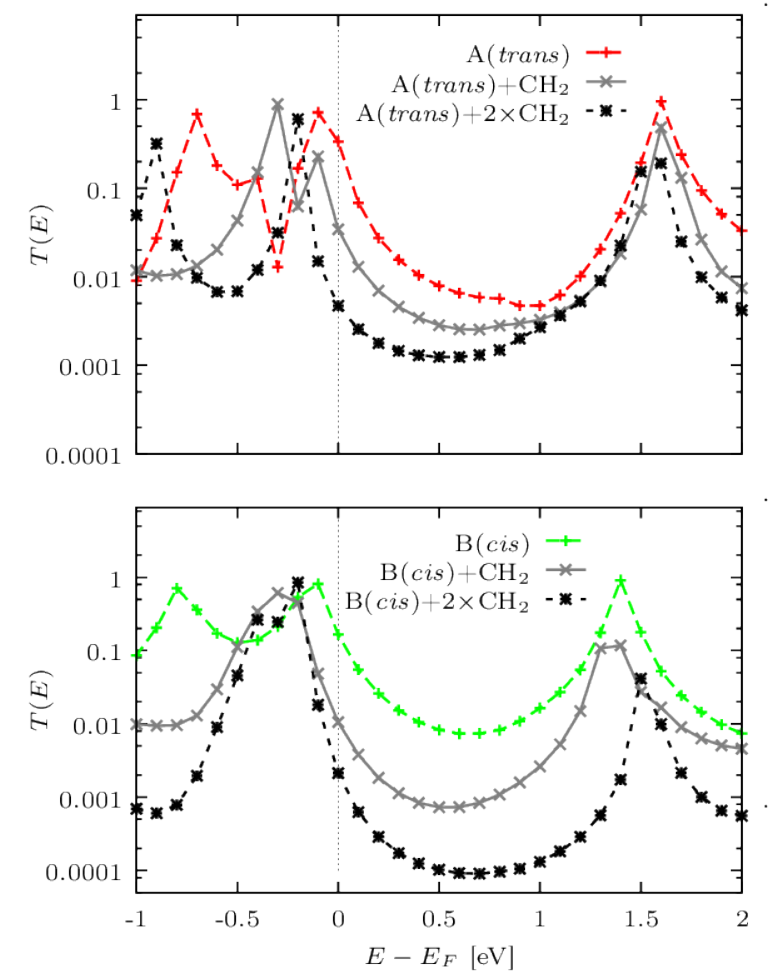}
\caption{
Comparison of transmissions of A(\textit{trans}) [red lines] and B(\textit{cis}) [green lines] junction conformations with one [full line], two [dotted line], and without [dashed line] methylene linker group(s) added to the molecule. Note the drop of the conductance by one order of magnitude for both conformations upon insertion of each of the methylene groups.
}
\label{fig:transmissionMe}
\end{figure}

Our results indicating strong rigidity of the conductance of DAB-based switch are at variance with the results of Zhang \textit{et al.}~\cite{zhang_04,zhang_06} who find that the conductance of the anchored \textit{trans} isomer is two orders of magnitude larger than that of the \textit{cis}. However, there is one major difference between our model junctions A, B, and C considered so far and the junction considered in Refs.~\cite{zhang_04,zhang_06,kim2012_1}, namely use of the anchoring methylene -CH$_2$- groups added at the ends of the AB molecule. This is important as presence of the methylene groups was made responsible for the small SR found also experimentally~\cite{kim2012_1}. The importance of the methylene groups on the transport has been experimentally studied on related systems by Danilov \textit{et. al.}~\cite{danilov_08}. Their results show that inclusion of methylene groups at both ends of the molecule changes the quantum transport regime from coherent, characterized by higher conductances, to the CB regime. In such a case, applying a small bias voltage the junction can be opened for the current, with the resulting differential conductance being 2-4 orders of magnitude smaller compared to the molecules without the methylene groups.
 
\begin{table}[!ht]
\begin{center}
\begin{tabular}{l|c|c}
\hline
\hline
$T(E_F)$                     &  A(\textit{trans}) & B(\textit{cis}) \\
\hline
Au-S-AB-S-Au                 & 0.274	         & 0.154             \\
Au-S-AB-CH$_2$-S-Au          & 0.034             & 0.0096            \\
Au-S-CH$_2$-AB-CH$_2$-S-Au   & 0.0047            & 0.0024            \\
\hline
\hline
\end{tabular}
\caption{
Comparison of transmissions of A(\textit{trans}) and B(\textit{cis}) junction conformations without, with a single/couple of methylene linker group(s) added to one/both side(s) of AB molecule. Note the drop of the conductance by one/two order(s) of magnitude for both conformations upon insertion of one/two methylene linker group(s), in qualitative agreement with experiment on related systems~\cite{danilov_08}. 
} 
\label{tab:methyl-G}
\end{center}
\end{table}

We have studied the role of the additional methylene groups for the A(\textit{trans}), and B(\textit{cis}) junctions, see Fig.~\ref{fig:geometry-Me}. Our model junctions A and B with two added methylene groups are now equivalent to those considered theoretically by Zhang \textit{et al.} in Refs.~\cite{zhang_04,zhang_06} and also to the junctions used experimentally~\cite{kim2012_1}. While addition of single methylene group on one side still leads to a molecule well coupled to one of the electrodes through one remaining thiol group, adding a methylene group on both sides of the AB molecule effectively separates it from the electrodes. This separation is expected to drive the nano-device into CB regime. 
Transmissions of the junctions modified by one and two additional methylene group(s) are compared to those of the unaltered junctions for A(\textit{trans}) and B(\textit{cis}) conformations of the junction in Fig.~\ref{fig:transmissionMe}. The most important change we observe is the decrease in the tunneling probability in the HOMO-LUMO (Lowest Unoccupied Molecular Level) gap, where a decrease by about on order of magnitude is observed after adding each of the two methylene groups. The effect is somewhat more pronounced in the \textit{cis} conformation. However, at the Fermi energy the decrease for both junctions is very similar, see Tab.~\ref{tab:methyl-G}, so that again no significant enhancement of the on/off ratio of the switching conductances is achieved. 

The magnitude of the decrease in conductance we find between unaltered junctions and those modified by the methylene linkers is in good agreement with the simple tunneling 
model through minimal \lq\lq poly-ethylene\rq\rq chain, where the transmission is exponentially dependent on the chain length via the expression~\cite{Tomfohr2002}
$$T \sim e^{-2 \beta l}~~,$$
where $\beta \approx 0.8$~\AA$^{-1}$ for -[CH$_2$]- chain~\cite{Ferretti2012}, and $l\approx 1.5$~\AA~we find for the C--C distance between the carbon atom on the methylene group and on a neighboring carbon on the AB molecule. Hence, we conclude that the role of the methylene linker is indeed to add a tunneling barrier between AB molecule and the sulfur atoms.

\begin{figure}[!h]
\includegraphics[width=0.5\textwidth]{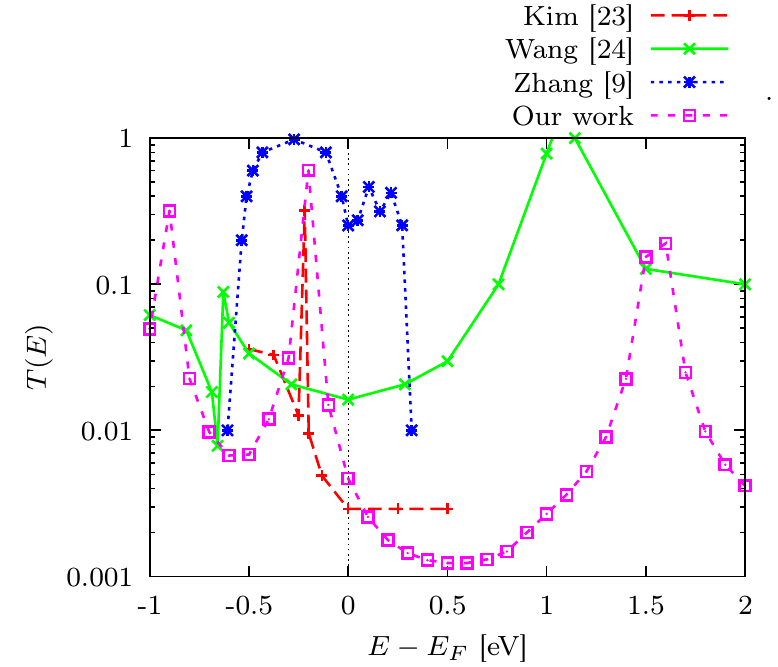}
\caption{
Comparison of transmissions for \textit{trans} conformation of the junction from our results for ABTM junction A(\textit{trans}) [pink line] with results of Kim \textit{et al.} [red line], Ref.~\cite{kim2012_1}, Wang and Cheng [green line], Ref.~\cite{wang2012}, and Zhang \textit{et al.} [blue line], Ref.~\cite{zhang_04}.
}
\label{fig:transmission-comparison}
\end{figure}

A comparison of our results with those obtained previously is shown in Fig.~\ref{fig:transmission-comparison}. From our results we conclude that even after modifying our junctions with methylene groups, which makes our junctions structurally very similar to those studied previously by Zhang \textit{et al.}~\cite{zhang_04,zhang_06}, our results are still both quantitatively \textit{and} qualitatively markedly different. Quantitatively, the conductance found in Refs.~\cite{zhang_04,zhang_06} for the \textit{trans} conformation with both methylene groups inserted is of similar magnitude to that we find for junctions without the methylene groups, hence about two orders of magnitude larger than we find for comparable junctions with the methylene groups, see Tab.~\ref{tab:methyl-G} and Fig.~\ref{fig:transmission-comparison}. Qualitatively their peak of the transmission is in similar position to our ($\approx$0.2 eV below $E_{F}$) but its width is substantially larger and there are additional features around $E_{F}$ leading to a very high \textit{trans} conductance and \textit{trans}/\textit{cis} conductance difference of two orders of magnitude not present in our model.
Comparing our results to results obtained by Wang and Cheng, who considered SAMs of ABTM~\cite{wang2012}, we again find conductance almost an order of magnitude smaller with peaks of transmission in different positions, $\sim$0.6 and 1.5 eV for \textit{trans} and $\sim$0.8 eV for \textit{cis} conformation, versus our values of $\sim$0.2 and 0.9 eV for \textit{trans} and $\sim$ 0.2 and 0.4 \textit{cis}. These differences indicate that ABTM sandwiched in tip and layer Au electrodes are indeed quite different transport systems.   
Contrary, comparison with results of Kim~\textit{et al.}~\cite{kim2012_1}, who also consider Au tip electrodes shows an excellent qualitative agreement for both \textit{trans} (Fig.~\ref{fig:transmission-comparison}) and \textit{cis} conformations. The only difference is in the fact that we find conductance of the 
\textit{trans} conformation slightly larger than \textit{cis}, whereas Kim~\textit{et al.} find higer conductances for the \textit{cis}. However, we have shown that the conductance is strongly dependent on details of the molecular conformation, so that for other conformations not sampled by our very limited protocol the results can easily be reverted. However, the very small \textit{trans}/\textit{cis} switching ratios remain robust.

\section{\label{sec:concl}Discussion and Conclusions}

We have presented a study of a single molecule azobenzene switch suspended via thiolate links between realistic models of gold tips with the quest to understand the SR of this nano-system. We use essentially the same tools as in the previous theoretical studies~\cite{zhang_04,zhang_06,wang2012,kim2012_1}, a combination of transfer matrix technique and density functional theory. However, there are two key differences between our study and the previous modeling, namely 1) use of realistic tip models mimicking the MCBJ set-ups, and 2) implementation of the theory in plane-wave basis set. In addition, we also systematically study the influence of the methylene linkers on transport properties. All these ingredients in our treatment are shown to have an important effect. Our MCBJ tip set-up provide results significantly different from set-up corresponding to SAMs-type of electrodes~\cite{wang2012}. The role of methylene side linkers as tunneling barriers each reducing the conductance by about an order-of-magnitude is fully borne out by our calculated conductances. In addition, compared to the previous calculations using localized basis sets, our treatment of the system in plane-wave basis is expected to result in a better description of the electrons in the metal electrodes. We find broadly similar conductances of both conformations, differing by a factor of at most $\approx$2. Most importantly, using our realistic tip models, we have been able to identify two  \textit{trans} conformations of the junction also showing conductance differences by about the same factor of $\approx$2, confirming the role played by the electrode/molecule geometry. Low SR and a large spread in results for the \textit{trans} conformation of the junction are in agreement with the experimental results and with modeling using tip electrodes~\cite{kim2012_1} and in line with results found for related systems~\cite{Quek2009} and in complete disagreement with results of Refs.~\cite{zhang_04, zhang_06}, who also use tip electrodes in their model. In qualitative agreement with experiments, we find that the same nano-device with one/two methylene linker group(s) inserted on one/both ends of the azobenzene molecule is driven into tunneling regime and reduces the conductances by up to two orders of magnitude, again almost uniformly for both conformations. This finding provides an argument against -CH$_2$- side chains being responsible for the small SR of the azobenzene-based molecular switch. Hence, present results make us conclude, that the only way to bring about large SR with AB/DAB molecule bridging gold electrodes appears to be by a complete breakage of the junction by applying pulling force as ductile gold electrodes are capable of breaking/remaking the junction in a reversible way~\cite{turansky_10_1,turansky_10_2}.

The reason for the large disagreement found with respect to results of Zhang \textit{et al.}~\cite{zhang_04, zhang_06} remains unclear. The reasons for a large conductance difference between the different isomers could be due to: a) big difference between the electronic structure of the two isomers, b) significant change of the Fermi level position with respect to the HOMO for the two isomers, or c) significant changes of the geometries of the junction depending on the isomer.
Our present study as well as our recent many-body study of the AB molecule~\cite{dubecky_10,dubecky_11} do not support reason a). The many-body study revealed deviations from experiments by as much as $\approx$1 eV with DFT techniques and even with medium-quality quantum chemistry methods~\cite{dubecky_10,dubecky_11}, depending on the level of correlation and/or basis set used. However, these differences affect mostly the singlet/triplet excited states of the molecule~\cite{dubecky_11}, rather than the states around the Fermi level responsible for the conductance of the contacted molecule which are very similar for both isomers. Reason b) is also not supported by our study as the Fermi level is pinned to the tail of the HOMO in all the systems studied. Reason c) appears the most probable case as we do see conductance dependence on the geometrical details of the junction. However, at least within the limited search conducted here, such changes can only account for much smaller conductance changes than those reported previously. Much smaller conductance differences of the order of~$\approx$ 10\% could also be explained by the differences in treatment of our electrodes and k-point sampling, more modern exchange-correlation functionals etc., see Sect.~\ref{sec:simulations}. While all of our approximations appear to be an improvement over the alternative previous choices, differences of two orders of magnitude for the \textit{trans} conformation are far too large to be accounted for by such technical issues. Ultimately the theoretical results must be benchmarked experimentally. Previous results suggesting huge SR appear to disagree with the recent experiment supporting low SR~\cite{kim2012_1} as well as with the trends imposed by -CH$_2$- acting as tunneling barriers~\cite{danilov_08}. 
All our findings provide a solid support to our conclusion of the \textit{rigidity of conductance} of the gold-DAB-gold molecular junction upon reisomerization and make us conclude that the conductance difference is far too small to be used as a base for conductance change in a molecular switch.

\begin{acknowledgements}
Work supported by APVV-0207-11, VEGA (2/0007/12) projects. This research was supported in part also by 
the European Community's FP7 grant under agreement no. 211956 as an outcome from the ETSF User Project No. 220
and by ERDF OP R \& D, project CE meta-QUTE ITMS 26240120022, and via CE SAS QUTE. 
The authors wish to acknowledge R{\'o}bert Turansk{\'y} for providing us with the geometries of the relaxed 
Au(110)-DAB-Au(110) nano-junctions.
\end{acknowledgements}


\begin{thebibliography}{10}

\bibitem{tans_97}
S.~J. Tans, M.~H. Devoret, H. Dai, T. A., R.~E. Smalley, L.~J. Geerligs, and C.
  Dekker, Nature (London) {\bf 386},  474  (1997).

\bibitem{reed_97}
M.~A. Reed, C. Zhou, C.~J. Muller, T.~P. Burgin, and J.~M. Tour, Science {\bf
  278},  252  (1997).

\bibitem{joachim_95}
C. Joachim, J.~K. Gimzewski, R.~R. Schlittler, and C. Chavy, Phys. Rev. Lett.
  {\bf 74},  2102  (1995).

\bibitem{moresco_95}
F. Moresco, G. Meyer, K.-H. Rieder, H. Tang, A. Gourdon, and C. Joachim, Phys.
  Rev. Lett. {\bf 86},  672  (2001).

\bibitem{emberly_03}
E.~G. Emberly and G. Kirczenow, Phys. Rev. Lett. {\bf 91},  188301  (2003).

\bibitem{li_04}
J. Li, G. Speyer, and O. F. Sankey, Phys. Rev. Lett. {\bf 93},  248302  (2004).

\bibitem{zhang_04}
C. Zhang, M.~H. Du, H.~P. Cheng, X.~G. Zhang, A.~E. Roitberg, and J.~L. Krause,
  Phys. Rev. Lett. {\bf 92},  158301  (2004).

\bibitem{zhang_06}
C. Zhang, Y. He, H.~P. Cheng, Y.~Q. Xue, M.~A. Ratner, X.~G. Zhang, and P.
  Krstic, Phys. Rev. B {\bf 73},  125445  (2006).

\bibitem{turansky_10_1}
R. Turansk\'{y}, M. Kon\^{o}pka, N.~L. Doltsinis, I. \v{S}tich, and D. Marx,
  ChemPhysChem {\bf 11},  345  (2010).

\bibitem{turansky_10_2}
R. Turansk\'{y}, M. Kon\^{o}pka, N.~L. Doltsinis, I. \v{S}tich, and D. Marx,
  Phys. Chem. Chem. Phys. {\bf 12},  13922  (2010).

\bibitem{qiu_04}
X.~H. Qiu, G.~V. Nazin, and W. Ho, Phys. Rev. Lett. {\bf 93},  196806  (2004).

\bibitem{choi_06}
B.-Y. Choi, S.-J. Kahng, S. Kim, H. Kim, H.~W. Kim, Y.~J. Song, J. Ihm, and Y.
  Kuk, Phys. Rev. Lett. {\bf 96},  156106  (2006).

\bibitem{milosevic_07}
V. Simic-Milosevic, M. Mehlhorn, K.-H. Rieder, J. Meyer, and K. Morgenstern,
  Phys. Rev. Lett. {\bf 98},  116102  (2007).

\bibitem{Comstock2005}
M. J. Comstock, Jongweon Cho, A. Kirakosian, and M. F. Crommie, Phys. Rev. B {\bf 72}, 153414 (2005).

\bibitem{Henningsen2008}
N.~Henningsen, R.~Rurali, K.~J.~Franke, I.~Fernández-Torrente, and J.~I.~Pascual, Applied Physics A {\bf 93}, 
241 (2008).

\bibitem{konopka_08}
M. Kon\^{o}pka, R. Turansk\'{y}, J. Reichert, H. Fuchs, D. Marx, and I.
  \v{S}tich, Phys. Rev. Lett. {\bf 100},  115503  (2008).

\bibitem{Quek2009}
S.~Y. Quek, M. Kamenetska, M.~L. Steigerwald, H.~J. Choi, S.~G. Louie, M.~S.
  Hybertsen, J.~B. Neaton, and L. Venkataraman, Nature Nanotechnology {\bf 4},
  230  (2009).

\bibitem{balzani_00}
V. Balzani, A. Credi, F.~M. Raymo, and J.~F. Stoddart, Angew. Chem. Int. Ed.
  {\bf 39},  3348  (2000).

\bibitem{feringa_00}
B.~L. Feringa, R.~A. van Delden, N. Koumura, and E.~M. Geertsema, Chem. Rev.
  {\bf 100},  1789  (2000).

\bibitem{collin_98}
J.~P. Collin, P. Gavi\~{n}a, V. Heitz, and J.~P. Sauvage, Eur. J. Inorg. Chem.
  {\bf 1},  1  (1998).

\bibitem{dulic_03}
D. Duli\'c, S.~J. van~der Molen, T. Kudernac, H.~T. Jonkman, J. J. D. de~Jong, T. N. Bowden, J. van Esch, B. L. Feringa, and B. J. van Wees, Phys. Rev. Lett. {\bf 91},  207402  (2003).

\bibitem{kim2012_2}
Y. Kim, J. Hellmuth, D. Sysoiev, F. Pauly, T. Pietsch, J. Wolf, A. Erbe, T.
  Huhn, U. Groth, U.~E. Steiner, and E. Scheer, Nanolett. {\bf 12},  3736
  (2012).

\bibitem{kim2012_1}
Y. Kim, A. Garcia-Lekue, D. Sysoiev, T. Frederiksen, U. Groth, and E. Scheer,
  Phys. Rev. Lett. {\bf 109},  226801  (2012).

\bibitem{wang2012}
Y. Wang and H.-P. Cheng, Phys. Rev. B {\bf 86},  035444  (2012).

\bibitem{afm-mcb-rev}
J.~M. van Ruitenbeek, A. Alvarez, I. Pi\~{n}eyro, C. Grahmann, P. Joyez, M.~H.
  Devoret, D. Esteve, and C. Urbina, Rev. Sci. Instrum. {\bf 67},  108  (1996).

\bibitem{Nozaki09}
D. Nozaki and G. Cuniberti, Nano Research {\bf 2},  648  (2009).

\bibitem{Valle2007}
M. del Valle, R. Gutierrez, C. Tejedor, and G. Cuniberti, Nature Nanotechnology
  {\bf 2},  176  (2007).

\bibitem{hugel_02}
T. Hugel, N.~B. Holland, A. Cattani, L. Moroder, M. Seitz, and H.~E. Gaub,
  Science {\bf 296},  1103  (2002).

\bibitem{buttiker_86}
M. Buttiker, Phys. Rev. Lett. {\bf 57},  1761  (1986).

\bibitem{pwcond}
A. Smogunov, A. Dal~Corso, and E. Tosatti, Phys. Rev. B {\bf 70},  045417
  (2004).

\bibitem{Strange08}
M. Strange, I.~S. Kristensen, K.~S. Thygesen, and K.~W. Jacobsen, J. Chem.
  Phys. {\bf 128},  114714  (2008).

\bibitem{Garcia-Gil2009}
S. Garci­a-Gil, A. Garcia and N. Lorente and P. Ordejon, Phys. Rev. B {\bf 79}, 075441 (2009).

\bibitem{danilov_08}
A. Danilov, S. Kubatkin, S. Kafanov, P. Hedeg{\aa}rd, N. Stuhr-Hansen, K.
  Moth-Poulsen, and T. Bj{\o}rnholm, Nano Letters {\bf 8},  1  (2008).

\bibitem{Espresso}
P. Giannozzi, S. Baroni, N. Bonini, M. Calandra, R. Car, C. Cavazzoni, D.
  Ceresoli, G.~L. Chiarotti, M. Cococcioni, I. Dabo, A. Dal~Corso, S. Fabris,
  G. Fratesi, S. de~Gironcoli, R. Gebauer, U. Gerstmann, C. Gougoussis, A.
  Kokalj, M. Lazzeri, L. Martin-Samos, N. Marzari, F. Mauri, R. Mazzarello, S.
  Paolini, A. Pasquarello, L. Paulatto, C. Sbraccia, S. Scandolo, G. Sclauzero,
  S.~A. P., A. Smogunov, P. Umari, and R.~M. Wentzcovitch, J. Phys. Condens.
  Matter {\bf 21},  395502  (2009), http:/www.quantum-espresso.org/.

\bibitem{vaderbilt_90}
D. Vanderbilt, Phys. Rev. B {\bf 41},  7892  (1990).

\bibitem{monkhorst_76}
H.~J. Monkhorst and J.~D. Pack, Phys. Rev. B {\bf 13},  5188  (1976).

\bibitem{pbe_1}
J.~P. Perdew, K. Burke, and M. Ernzerhof, Phys. Rev. Lett. {\bf 77},  3865
  (1996).

\bibitem{pbe_2}
J.~P. Perdew, K. Burke, and M. Ernzerhof, Phys. Rev. Lett. {\bf 78},  1396
  (1997), errata.

\bibitem{Joon1999}
H. J. Choi and J. Ihm, Phys. Rev. B {\bf 59},  2267  (1999).

\bibitem{Koentopp05}
M. Koentopp, K. Burke, and F. Evers, Phys. Rev. B {\bf 73},  121403(R)  (2006).

\bibitem{Ferretti05}
A. Ferretti, A. Calzolari, R.~Di Felice, F. Manghi, M.~J. Caldas, M. Buongiorno
  Nardelli, and E. Molinari, Phys. Rev. Lett. {\bf 94},  116802  (2005).

\bibitem{Quek07}
S.~Y. Quek, L. Venkataraman, H.~J. Choi, S.~G. Louie, M.~S. Hybertsen, and
  J.~B. Neaton, Nano Letters {\bf 7},  3477  (2008).

\bibitem{Toher08}
C. Toher and S. Sanvito, Phys. Rev. B {\bf 77},  155402  (2008).

\bibitem{Mera10}
H. Mera and Y.~M. Niquet, Phys. Rev. Lett. {\bf 105},  216408  (2010).

\bibitem{Kurth10}
S. Kurth, G. Stefanucci, E. Khosravi, C. Verdozzi, and E.~K.~U. Gross, Phys.
  Rev. Lett. {\bf 104},  236801  (2010).

\bibitem{Thygesen09}
K.~S. Thygesen and A. Rubio, Phys. Rev. Lett. {\bf 102},  046802  (2009).

\bibitem{Stadler2008}
R. Stadler, V. Geskin, and J. Cornil, Phys. Rev. B {\bf 78},  113402  (2008).

\bibitem{Zotti2010}
L.~A. Zotti, T. Kirchner, J.-C. Cuevas, F. Pauly, T. Huhn, E. Scheer, and A.
  Erbe, Small {\bf 6},  1529  (2010).

\bibitem{emt}
K.~W. Jacobsen, J.~K. Norskov, and M.~J. Puska, Phys. Rev. B {\bf 35},  7423
  (1987).

\bibitem{Heimel2006}
G. Heimel, L. Romaner, J.-L. Bredas, and E. Zojer, Phys. Rev. Lett. {\bf 96},
  196806  (2006).

\bibitem{Tomfohr2002}
J.~K. Tomfohr and O.~F. Sankey, Phys. Rev. B {\bf 65},  245105  (2002).

\bibitem{Ferretti2012}
A. Ferretti, G. Mallia, L. Martin-Samos, G. Bussi, A. Ruini, B. Montanari, and
  N.~M. Harrison, Phys. Rev. B {\bf 85},  235105  (2012).

\bibitem{dubecky_10}
M. Dubeck{\'y}, R. Derian, L. Mitas, and I. {\v S}tich, J. Chem. Phys. {\bf
  133},  244301  (2010).

\bibitem{dubecky_11}
M. Dubeck{\' y}, R. Derian, L. Horv{\'a}thov{\'a}, M. Allan, and I. {\v S}tich,
  Phys. Chem. Chem. Phys. {\bf 13},  244301  (2011).

\end{thebibliography}

\end{document}